\documentclass[
    ,final            
  ]
  {aipproc}

\layoutstyle{8x11single}

\newcommand{\gev}{\,{\rm GeV}}

\begin{document}

\title{Probing GPDs in Ultraperipheral Collisions}


\author{D.Yu. Ivanov}{
address = {Sobolev Institute of Mathematics and Novosibirsk State University,630090 Novosibirsk, Russia}
}

\author{B. Pire}{
  address={CPHT, {\'E}cole Polytechnique, CNRS, 91128 Palaiseau, France}
}

\author{L. Szymanowski}{
  address={National Centre for Nuclear Research (NCBJ), Warsaw, Poland}
}

\author{J. Wagner}{
  address={National Centre for Nuclear Research (NCBJ), Warsaw, Poland}
}

\begin{abstract}
Ultraperipheral collisions in hadron colliders give new opportunites to investigate the hadron stucture through exclusive photoproduction processes. We describe the possibility of measuring the Generalized Parton Distributions in the Timelike Compton Scattering process and in the production of heavy vector meson.
\end{abstract}

\maketitle


\section{Introduction}

Besides their primary use for exploring a new energy domain, high energy hadron colliders are powerful sources of quasi real photons in ultraperipheral collisions \cite{UPC}. This is usually described through the equivalent photon approximation (EPA) formula that reads
\begin{eqnarray}
\sigma^{AB} = 
\int dk_A
\frac{dn^A}{dk_A}\sigma^{\gamma B}(W_A(k_A)) + 
 \int dk_B
\frac{dn^B}{dk_B}\sigma^{\gamma A}(W_B(k_B)) \nonumber
\end{eqnarray}
where $k_{A,B} = \frac{1}{2}x_{A,B}\sqrt{s}$. and $\frac{dn}{dk}$ is an equivalent photon flux, i.e. the number of photons with energy $k$.

This opens the possibility for  studying many aspects of photon proton, photon nucleus and photon photon collisions at ultra high energies,  particularly at the LHC, many years before the eventual construction of electron-ion colliders \cite{EIC}. 
The high luminosity and energies of these quasi-real photon beams open a new kinematical domain for the study of exclusive processes which are now understood in the framework of the colinear factorization approach of QCD, as a powerful tool to our understanding of how quarks and gluons build hadrons.  The concept of generalized parton distributions (GPDs)\cite{gpdrev} is central in this respect. In particular the transverse location of quarks and gluons become experimentally measurable via the transverse momentum dependence of the GPDs \cite{Burk}. Determining sea-quark and gluon GPDs in the small skewedness region is an essential program complementary to the determination of the valence quark GPDs at lower energy electron accelerators.

The golden channel to access GPDs in quasi real photon processes is lepton pair production with a large invariant mass Q, either in the continuum or near a charmonium resonance such as $J/\Psi$. In the continuum, the process known as timelike Compton scattering (TCS) \cite{TCS} is the timelike analog of the celebrated deeply virtual Compton scattering (DVCS) which has been and is the subject of intense studies at medium and high energies. $J/\Psi$ production has the advantage of larger cross sections but may depend on the way the charmonium wave function is described. Both processes probe the same underlying partonic dynamics.

\section{Timelike compton Scattering}
The proof of QCD colinear factorization of TCS at leading twist follows the same line as the one for DVCS. This solidly establishes the validity of the approach.
As in the case of DVCS, a purely electromagnetic competing mechanism, the Bethe-Heitler (BH) mechanism contributes at the amplitude level to the same final state as TCS. This BH process has a very peculiar angular dependence and overdominates the TCS process if one blindly integrates over the final phase space. This is the reason why most Monte Carlo programs for ultraperipheral collisions do not consider the QCD process we are discussing here. A winning strategy consists of choosing kinematics where the amplitudes of the two processes are of the same order of magnitude, and in using specific observables sensitive to the interference of the two amplitudes.

We estimated \cite{TCSUPC} the Born order lepton pair production in UPC at LHC, including both processes with various cuts to enable the study of the TCS contribution.
The pure Bethe - Heitler contribution to $\sigma_{p p}$, integrated over  $\theta = [\pi/4,3\pi/4]$, $\phi = [0,2\pi]$, $t =[-0.05 \gev^2,-0.25 \gev^2]$, ${Q'}^2 =[4.5 \gev^2,5.5 \gev^2]$, and photon energies $k =[20,900]\gev $  gives:
\begin{equation}
\sigma_{pp}^{BH} = 2.9 \mbox{pb} \;.
\nonumber
\end{equation}  
The leading order Compton contribution (calculated with NLO GRVGJR2008 PDFs, and $\mu_F^2 = 5 \gev^2$) gives:
\begin{equation}
\sigma_{pp}^{TCS} = 1.9 \mbox{pb}\;. 
\nonumber
\end{equation}
We have choosen the range of photon energies in accordance with expected capabilities to tag photon energies
at the LHC. This amounts to a large rate of order of $\sim 10^5$ events/year at the LHC with nominal 
luminosity ($10^{34}\,$cm$^{-2}$s$^{-1}$). 

It should be stressed that  the crossing from a spacelike to a timelike probe is an important test of the understanding of QCD corrections, as shown by the history of the understanding of inclusive processes in terms of QCD. In the case of exclusive processes, the difference between coefficient functions in the timelike vs spacelike regimes can be traced back to the analytic structure (in the $q^2$ variable) of the scattering amplitude \cite{MPSW}. We found that $O(\alpha_s)$ corrections are rather large in timelike processes \cite{PSW,MPSSW} and quite factorization scale dependent. Moreover they depend strongly on the gluon GPDs, which  is the signature of a powerful tool to extract hadronic information from experimental data.

\section{Heavy Vector Meson Production}
The photoproduction of the heavy vector meson:
\begin{equation}
\gamma p \to V p
\end{equation}
is a subject of intense experimental \cite{VMexp} and theoretical \cite{VMth} studies. The main motivation of such studies is the possibility to explore gluon densities in the nucleon. In this work we present preliminary results \cite{Prep} on the use of the collinear factorization approach at the next to leading order in $\alpha_S$, which was developed in \cite{HVMP} , in the context of ultraperipheral collisons.

\begin{figure}
\includegraphics[keepaspectratio,width=0.45\textwidth,angle=0]{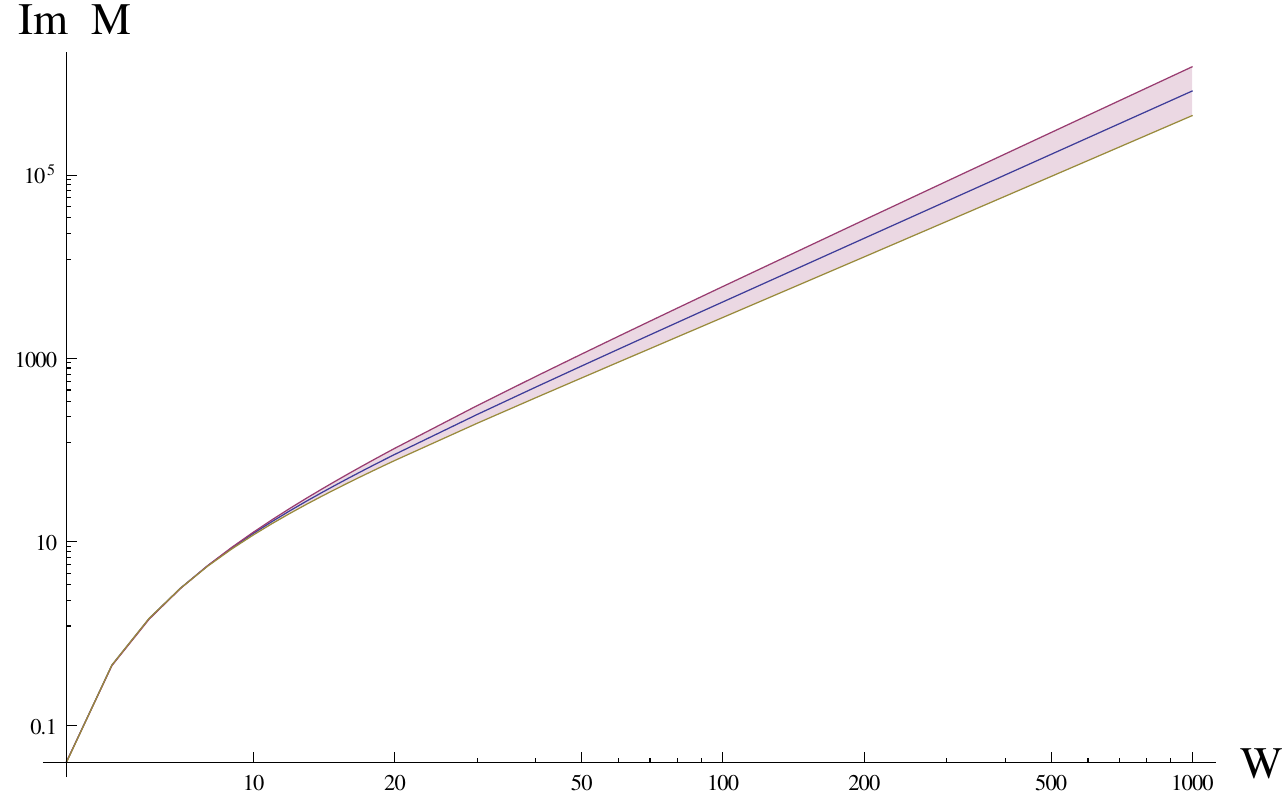}
\includegraphics[keepaspectratio,width=0.45\textwidth,angle=0]{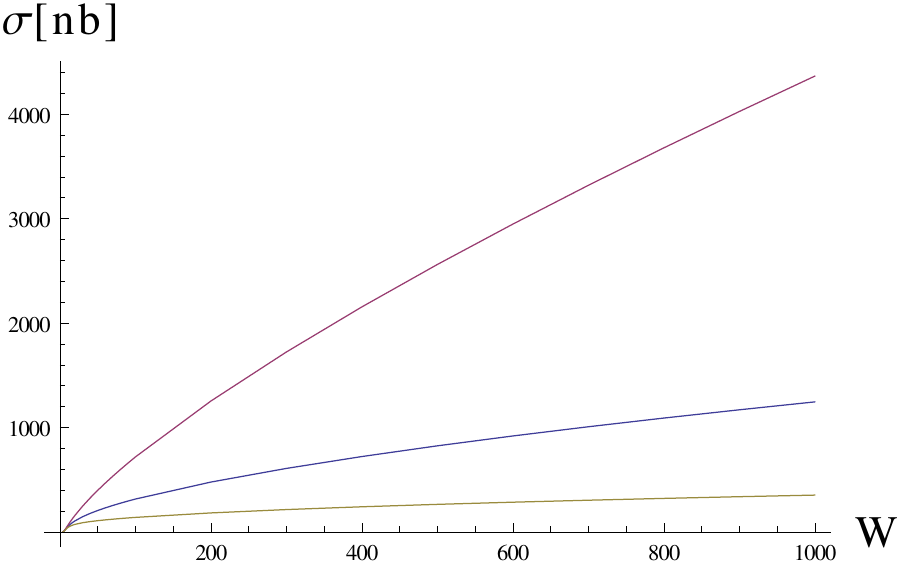}
\caption{(left) Imaginary part of the amplitude $\mathcal M$ and (right) cross section for photoproduction of $J/\psi$ at the LO as a function of $W = \sqrt{s_{\gamma p}}$ for $\mu_F^2 = M_{J/\psi}^2 \times \{0.5,1,2\}$ (respectively from bottom to top).
} 
\label{fig:JPsi_LO}
\end{figure}

The amplitude $\mathcal{M}$ is given by factorization formula:
\begin{eqnarray}
{\cal M}
&\sim &
\left(\frac{\langle O_1 \rangle_V}{m^3}\right)^{1/2}
 \int\limits^1_{-1} dx
\left[\, T_g( x,\xi)\, F^g(x,\xi,t)+
T_q (x,\xi) F^{q,S} (x,\xi,t) \, 
\right] \, ,
\\ 
F^{q,S} (x,\xi,t)&=&\sum_{q=u,d,s}  F^q (x,\xi,t) \, .
\end{eqnarray}
where $m$ is the pole mass of the heavy quark,  $\langle O_1 \rangle_V$ is given by the NRQCD through leptonic meson decay rate, $\xi$ is the fraction of the longitudinal momentum transfer, $t$ is the momentum transfer squared, $F^{q,g}$ are quark and gluon GPDs and $T^{q,g}$ are hard scattering coefficient functions given by:

%

\begin{equation}
 T_g(x,\xi)=\frac{\xi}{(x-\xi+i\varepsilon)(x+\xi-i\varepsilon)}
{\cal A}_g\left(\frac{x-\xi+i\varepsilon}{2\xi}\right) \quad , \quad
T_q( x,\xi)={\cal A}_q\left(\frac{x-\xi+i\varepsilon}{2\xi}\right) \, .
\label{gAT}
\end{equation}
At the leading order they read:
\begin{equation}
{\cal A}_g^{(0)}(y)=\alpha_S \quad , \quad
{\cal A}_q^{(0)}(y)=0 \, .
\label{LOq}
\end{equation}
In the paper \cite{HVMP} the leading order result for ${\cal A}_g^{(0)}(y)$ contains a mistake\footnote{Authors would like to thank Stephen Jones for pointing out this mistake} due to the incorrect treatment of the gluon polarizations in D dimensions. It is a wrong factor $D-4$ in the leading order that makes some final influence on the next to leading order results for the quark NLO  coefficient function: namely in the first line of eq.3.70 of \cite{HVMP} one should change $\left(\log\frac{4m^2}{\mu_F^2} -1 \right) \rightarrow \left(\log\frac{4m^2}{\mu_F^2} \right)$ \cite{erratum}. Despite this difference, the main phenomenological results remain the same: the NLO corrections are very big and the overall result depends very strongly on the choice of the factorization scale. In Fig.\ref{fig:JPsi_LO} we present the leading order result for $J/\psi$ photoproduction using the Goloskokov-Kroll model of GPDs \cite{GK}. On the left hand side of this figure we show the imaginary part of the amplitude $\mathcal{M}$, plotted for the following values of factorization scale $\mu_F^2 = M_{J\psi}^2\cdot\{0.5,1,2\}$ (respectively from bottom to top). The factorization scale dependence of the LO result is visible even more dramatically on the right hand side of Fig.\ref{fig:JPsi_LO}, where the cross section is plotted. For the  high values of the center of mass energy $W$ the results for $\mu_F^2 = 0.5  M_{J\psi}^2$ and $\mu_F^2 = 2  M_{J\psi}^2$ differ by one order of magnitude.

\begin{figure}
\includegraphics[keepaspectratio,width=0.65\textwidth,angle=0]{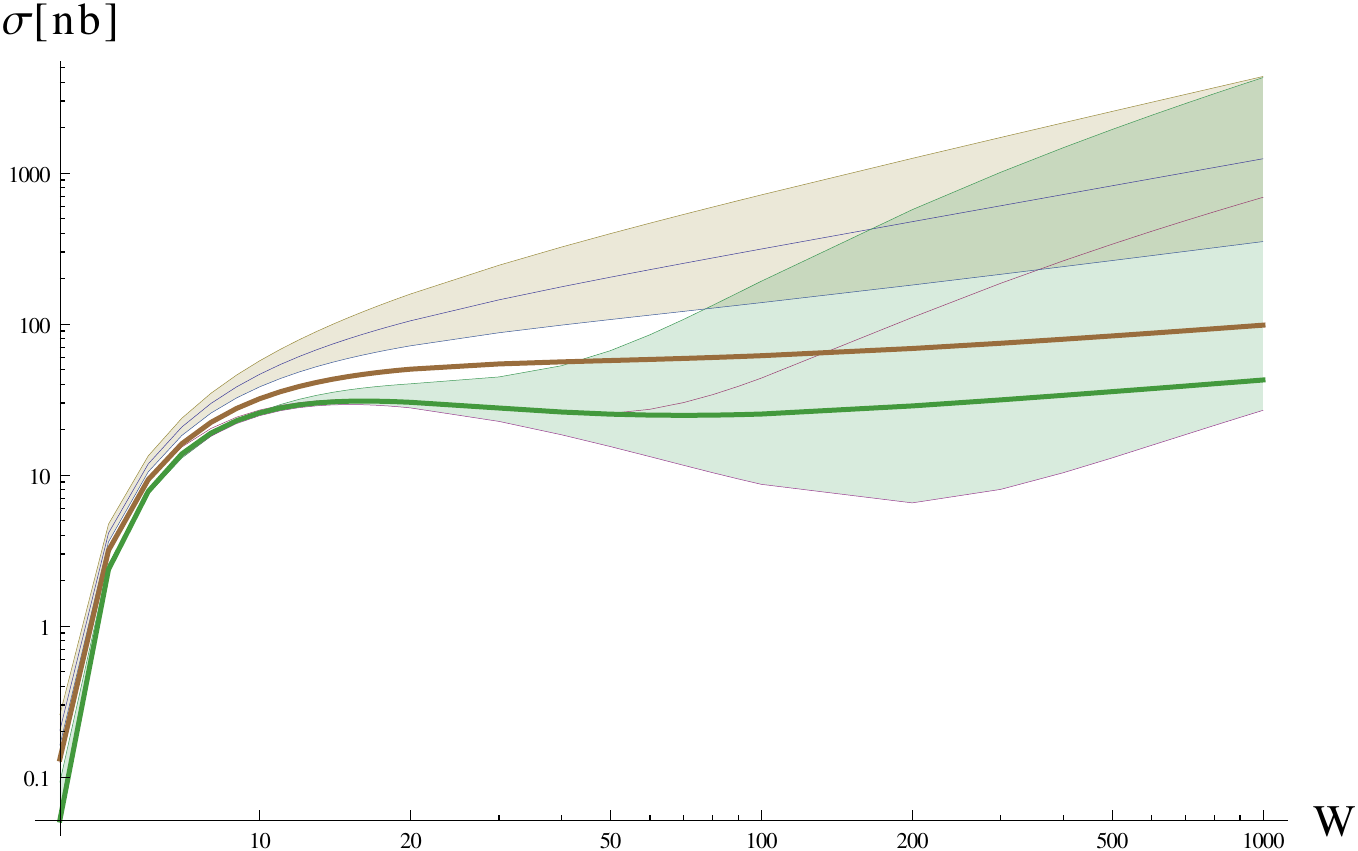}
\caption{$J/\psi$ photoproduction cross section as a function of $W = \sqrt{s_{\gamma p}}$ for $\mu_F^2 = M_{J/\psi}^2 \times \{0.5,1,2\}$- LO (purple band) and NLO (blue band). Thick lines for LO(red) and NLO(green) for $\mu_F^2 = 1/4  M_{J/\psi}^2 $. } 
\label{fig:JPsi_NLO}
\end{figure}

Results for the NLO cross section are presented on the fig.\ref{fig:JPsi_NLO}. Blue(purple) band presents the results for LO(NLO) cross section for the factorization scale in the range $\mu_F^2 = M_{J\psi}^2\cdot\{0.5,1,2\}$. Dependence on that scale is still very big for the high values of $W$. This behaviour is connected to the low-$\xi$ (high $W$) behaviour of the coefficient functions. In that regime the imaginary part of the amplitude dominates, and the relative size of the corrections reads: 
\begin{equation}
\sim \frac{\alpha_S(\mu_R)
N_c}{\pi}\ln\left(\frac{1}{\xi}\right)
\ln\left(\frac{\frac{1}{4}M_V^2}{\mu_F^2}\right)
\nonumber\ .
\label{ux}
\end{equation}
The size of the corrections, and the sensitivity of the NLO result to the factorization scale choice, shows that some additional information is needed to provide reliable theoretical predictions. This may come in terms of some scale fixing procedure \cite{BLM} or some artificial choice of the scale to minimalize the one-loop corrections (this possibility is shown on the Fig.\ref{fig:JPsi_NLO} - thick red (green) line corresponds to the LO (NLO) result for $\mu_F^2 = 1/4 M_V^2$) or on the resummation of the large terms \cite{Dima:Blois}.
%
%

\begin{figure}
\includegraphics[keepaspectratio,width=0.48\textwidth,angle=0]{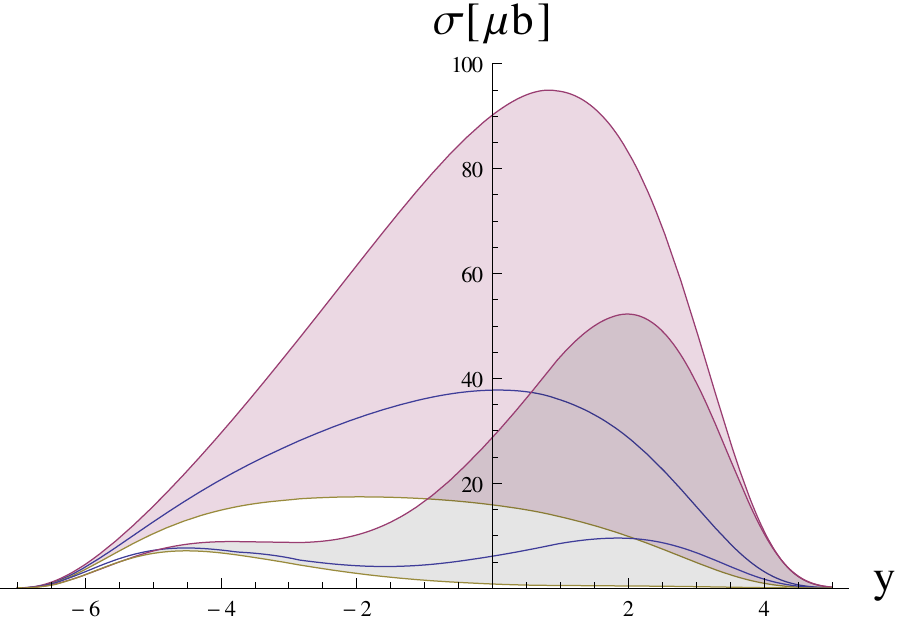}
\includegraphics[keepaspectratio,width=0.48\textwidth,angle=0]{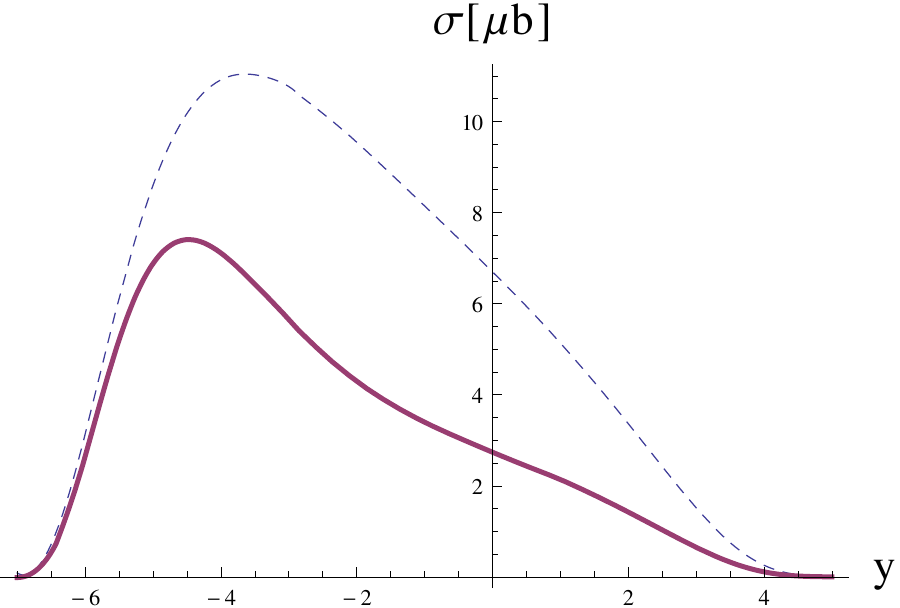}
\caption{$J/\psi$ photoproduction cross section in Ultraperipheral p-Pb collision, $\sqrt{s} = 5 \textrm{~TeV}$. 
(left) LO and NLO  $\mu_F^2 = M_{J/\psi}^2 \times \{0.5,1,2\}$. (right) LO and NLO for $\mu_F^2 = 1/4  M_{J/\psi}^2 $.
} 
\label{fig:UPC}
\end{figure}

%
%
%
Abovementioned problems influence of course the predictions for ultraperipheral collisions at the LHC. We illustrate it with the Fig.\ref{fig:UPC}, where we show the cross section for the photoproduction of the $J/\psi$ meson as a function of its rapidity, in the ultraperipheral p-Pb collision with $\sqrt{s} = 5 \textrm{~TeV}$.

\begin{theacknowledgments}
This work is partly supported by the Polish Grant NCN No DEC-2011/01/D/ST2/02069, by the COPIN-IN2P3 Agreement and by the grant RFBR-13-02-00695-a.
\end{theacknowledgments}



\bibliographystyle{aipproc}   



\end{document}